\documentclass{aa}
\usepackage[]{graphicx}
\usepackage[]{psfig}
\usepackage[]{epsfig}
\usepackage[]{longtable}
\usepackage{natbib}
\begin{document}
\newcommand{\feh}{[Fe/H]}
\newcommand{\Msun}{$M_{\odot}$}

\def\fdeg{\hbox{$.\mkern-4mu^\circ$}}
\def\farcs{\hbox{$.\!\!^{\prime\prime}$}}
\def\farcm{\hbox{$.\mkern-4mu^\prime$}}
\def\degr{\hbox{$^\circ$}}
\def\arcmin{\hbox{$^\prime$}}
\def\arcsec{\hbox{$^{\prime\prime}$}}
\def\sun{\hbox{$\odot$}}
\def\logg{\hbox{$\log g$}}
\def\Teff{\hbox{$T_{\rm eff}$}}
\def\fh{\hbox{$.\!\!^{\rm h}$}}
\def\fm{\hbox{$.\!\!^{\rm m}$}}
\def\fs{\hbox{$.\!\!^{\rm s}$}}
\def\hou{^{\rm h}}
\def\min{^{\rm m}}
\def\ssec{^{\rm s}}

\def\logg{log\hspace*{1mm}$g$}
\def\mfe{$\langle{\rm Fe}\rangle$}
\def\mh{$\langle{\rm H}\rangle$}
\def\mca{$\langle{\rm Ca}\rangle$}
\def\mmg{$\langle{\rm Mg}\rangle$}
\def\dmg{$\overline{\delta\langle{\rm Mg}\rangle}$}
\def\dca{$\overline{\delta\langle{\rm Ca}\rangle}$}
\def\dcn{$\overline{\delta S3839}$}
\def\dch{$\overline{\delta CH4300}$}
\def\dcnch{$\overline{\delta\langle{\rm CN+CH}\rangle}$}

\def\unit#1{\;{\rm #1}}


\def\wcen{$\omega$\,Cen}
\def\wcentauri{$\omega$\,Centauri}

\def\apj{ApJ}
\def\aap{A\&A}
\def\aaps{A\&AS}
\def\araa {RAA\&A}
\def\aj{AJ}
\def\apjl{ApJL}
\def\apjs{ApJS}
\def\asp{ASP}
\def\pasp{PASP}
\def\physrep{Phys.Rep.}
\def\apss{Ap\&SS}
\def\nat{Nature}
\def\mnras{MNRAS}


\title{Medium resolution spectroscopy in $\omega$ Centauri: \\ 
abundances of 400 subgiant and turn-off region stars\thanks{Based on 
observations obtained at the European Southern Observatory, Chile (Observing
Programme 69.D--0172).}}

\author {Andrea Kayser \inst{1,2} \and Michael Hilker \inst{1} \and Tom 
Richtler \inst{3} \and Philip G. Willemsen \inst{1}}

\offprints {A.~Kayser}
\mail{akayser@astro.unibas.ch}

\institute{
Sternwarte der Universit{\"a}t Bonn, Auf dem H{\"u}gel 71, 53121 Bonn, Germany
\and
Astronomisches Institut der Universit{\"a}t Basel, Venusstrasse 7, 4102 Binningen, Switzerland
\and
Universidad de Concepci{\'o}n, Departamento de F{\'\i}sica, Casilla 106-C,
Concepci{\'o}n, Chile
}

\date{Received / Accepted }

\titlerunning{Medium resolution spectroscopy in $\omega$ Centauri}

\authorrunning{A.~Kayser et al.}

\abstract{Medium resolution spectra of more than 400 subgiant and turn-off 
region stars in $\omega$ Centauri were analysed.
The observations were performed at the VLT/Paranal with FORS2/MXU.
In order to determine the metallicities of the sample stars, we defined a set of 
line indices (mostly iron) adjusted to the resolution of our spectra.
The indices as determined for \wcen\ were then compared to line indices from stars in the chemically homogeneous globular cluster M55, in addition to standard 
stars and synthetic spectra.
The uncertainties in the derived metallicities are of the order of $\pm 0.2\unit{dex}$.

Our study confirms the large variations in iron abundances found on the 
giant branch in earlier studies ($-2.2 <$[Fe/H]$< -0.7\unit{dex}$). 
In addition, we studied the 
$\alpha$-element and CN/CH abundances. Stars of different 
metallicity groups not only show distinct ages \citep{H/K:04}, but also 
different behaviours in their relative abundances. 
The $\alpha$ abundances increase smoothly with increasing metallicity resulting in a flat
[$\alpha$/Fe] ratio over the whole observed metallicity range. The combined
CN+CH abundance increases smoothly with increasing iron abundance. The most metal-rich stars
are CN-enriched.
In a CN vs. CH plot, though, the individual abundances divide into CN- and CH-rich branches.
The large abundance variations observed in our sample of (unevolved) subgiant branch stars most probably have their origin in the pre-enriched material rather than in internal
mixing effects. Together with the age spread of the different 
sub-populations, our findings favour the formation of $\omega$ 
Centauri within a more massive progenitor.
}

\maketitle



\section{Introduction}\label{intro}
In many respects, one of the most extraordinary globular clusters within 
our Milky Way is \wcentauri.
It is not only the most massive ($3\times 10^{6}\; M_{\odot}$) and 
with a projected ellipticity of 0.18 one of the most flattened clusters
in our galaxy, but it furthermore has a retrograde orbit around 
the Galactic centre, unlike most other globular clusters.

\begin{table*}[t]
\caption{\label{tab1}Log of observations}
\centering
\begin{tabular}[l]{clcll}
\hline\hline
 Date & Mask & Position (RA; DEC (J2000)) & Exp.time (1400V) & Exp.time (600I) \\
 \hline
  07/05/2002 & ome-1 & 201.950389 $\;$ -47.45774 & $3 \times 900 \unit{s}$ & $2 \times 900 \unit{s}$ \\
  \hline
  08/05/2002 & ome-2 & 201.950367 $\;$ -47.45791 & $3 \times 900 \unit{s}$ & $2 \times 1200 \unit{s}$ \\
  \hline
  09/05/2002 & ome-3 & 201.950467 $\;$ -47.45834 & $3 \times 900 \unit{s}$ & $3 \times 1200 \unit{s}$ \\
             & omw-1 & 201.499877 $\;$ -47.47863 & $3 \times 900 \unit{s}$ & $3 \times 1200 \unit{s}$ \\
   \hline
  10/05/2002 & omn-1 & 201.694769 $\;$ -47.31202 & $3 \times 900 \unit{s}$ & $3 \times 1200 \unit{s}$ \\
             & oms-1 & 201.740967 $\;$ -47.62819 & $3 \times 900 \unit{s}$ & $3 \times 1200 \unit{s}$ \\
             & omss-1 & 201.739449 $\;$ -47.72814 & $3 \times 900 \unit{s}$ & $3 \times 900 \unit{s}$ \\
             & M55-1 & 294.996455 $\;$ -30.88368 & $3 \times 600 \unit{s}$ & $3 \times 720 \unit{s}$ \\
  \hline
  11/05/2002 & oms-2 & 201.739332 $\;$ -47.62879 & $3 \times 720 \unit{s}$ & $3 \times 900 \unit{s}$ \\
             & omn-2 & 201.694567 $\;$ -47.31218 & $3 \times 900 \unit{s}$ & $3 \times 960 \unit{s}$ \\
             & omw-2 & 201.512185 $\;$ -47.48084 & $3 \times 720 \unit{s}$ & $3 \times 900 \unit{s}$ \\
  \hline
  12/05/2002 & omw-3 & 201.512222 $\;$ -47.48080 & $2 \times 900 \unit{s}$ & $2 \times 1200 \unit{s}$ \\
\hline\hline
\end{tabular}
\end{table*}

Although \wcen\ is one of the longest-known globular clusters and has been the 
subject of countless studies, the history of its investigation is more one of 
adding enigmatic properties than explaining them.
Already \citet{D/W:67} realised the unusual width of the red giant 
branch (RGB), which was later explained as a spread in metallicity by 
\citet{C/S:73} and which was spectroscopically confirmed by \cite{F/R:75}.
Further spectroscopic analyses revealed significant variations in nearly all element 
abundances \citep[e.g.][]{N/DC:95, S/S:00}.
Especially remarkable is the wide spread in [Fe/H], which covers a range of 
$-2.0$ to $-0.5\unit{dex}$. Both photometric and spectroscopic studies point 
to multiple stellar populations within \wcen , which differ not only in metal 
abundances but also in their spatial and kinematic properties.
The stars on the RGB in \wcen\ can be divided into at least 3 sub-populations. 
A dominant metal-poor component (peaking at [Fe/H]$\sim -1.7\unit{dex}$) 
contributes about 70\% of the stars. This population rotates at a maximum
velocity of $V_{\rm rot,max} \sim 8\unit{km \: s^{-1}}$ \citep{M/M/M:97}. 
In contrast, the 
intermediate metallicity population (about 25\% of the stars and peaking at 
[Fe/H]$\sim 
-1.2\unit{dex}$) shows no rotation but rather a concentration towards the cluster 
centre \citep{N/F:97}. 
The most metal-rich component comprises about 5\% of the stars with 
metallicities around $-0.6\unit{dex}$, and its spatial distribution is
off-centre with respect to the more metal-poor populations \citep{H/R:00}.
One has to be aware of the changes in nomenclature of these populations over the years.
Before the discovery of the most metal-rich population \citep{L/J:00, P/F:00}, what are
here called metal-poor and intermediate metallicity populations
are often referred to as metal-poor and metal-rich.

Recent high-resolution multi-band images reveal that the population puzzle
in \wcen\  is probably even more complicated.
 \citet{F/S:04}
discovered, apart from the main subgiant branch (SGB), an additional, narrow, and well-defined SGB in the colour-magnitude diagram (CMD) that seems to be the extension of the very red RGB
found by \citet{L/J:00} and \citet{P/F:00}.
Also, finer substructures have
been identified on the RGB and the main sequence (MS) using HST photometry.
\citet{S/F/P:05} have
revealed the subdivision of the intermediate metallicity population on the RGB
into at least 3 discrete populations.
Very stunning is also the bifurcation of the MS
\citep[e.g.][]{B/P:04}.
The observed fact that the red branch of the MS contains the majority of the
stars, which is the opposite of what is observed on the RGB, was at first very
puzzling. An unusual Helium enrichment of the intermediate metallicity population
now seems to be one of the most favoured explanations for this feature
\citep[e.g.][]{N:04, P/V:05}.

Both spectroscopic and broad and narrow band photometric studies
suggest a spread in age \citep[e.g.][]{L/J:00, H/W:00, H/R:00, R/L:04, H/W:04}.
All of these independent measurements reveal a significant age difference
between the sub-populations of 2-6 Gyr, correlated in the sense that
the younger populations are the more metal-rich ones.

Neither the spread in [Fe/H] nor in age has been observed in any other
globular cluster so far. Presumably, the formation history of \wcentauri\ differs
fundamentally from the ones of the other globular clusters in the Milky Way.
There are many different possibilities discussed as conceivable origins of the
variations in metal abundances and the observed age spread.
One of the most convincing ideas is that \wcen\ has an
extragalactic origin, either the isolated nucleus of a formerly disrupted
dwarf galaxy or a super-star cluster that has formed in an ancient, violent
merger event \citep[e.g.][]{B/F:03, F/K:03} .

Most of the spectroscopic studies of \wcen\ have so far been concentrated on the RGB.
These gave very accurate metallicity measurements, but only rough age estimates due to the
age-insensitivity of the location of RGB stars.
With pure photometric investigations, though, it is not possible to achieve
accurate metallicity determinations.
With 8-meter telescopes, however, it is now possible to obtain spectra of the
appropriate resolution of stars near the main sequence turn-off (MSTO) and the SGB, i.e. of stars in a region
that is highly sensitive to age.

In this paper, the results of an analysis of more than 400 spectra of stars in
this region of the CMD are presented. In particular, the method that has been
used for the Fe calibration is described in detail in this paper.
Section~\ref{obsred} introduces the observations and summarises the data
reduction. The selection of spectra for the final sample is described in
Sect.~\ref{grid}. Section~\ref{linemeas} deals with the definition of the
new line indices used to analyse this spectroscopic data.
A detailed description of the dependencies of the measured indices on stellar
parameters and the absolute calibration of the Fe indices is given in
Sects.~\ref{quali} and \ref{abscal}. In Sect.~\ref{other} we present a
qualitative analysis of other lines besides Fe, including Mg, Ca, CH, and CN.
A discussion and summary of the results is given in Sect.~\ref{summcon}.
The age spread of the different sub-populations in \wcen\ was presented in a
first paper by \citet{H/K:04}, and an automated analysis of a larger sample of stellar spectra
is presented in \citet{W/H:05}.


\section{Observations and data reduction}\label{obsred}
The instrument used for this project was the Focal Reducer and Low Dispersion
Spectrograph 2 (FORS2) attached to the Very Large Telescope (VLT/UT4) at
ESO/Paranal (Chile). FORS2 is equipped with a mask exchange unit (MXU) and
provides (with the standard resolution collimator) a field of view of
$6\farcm 8\times 6\farcm 8$.
The detector system consists of two $4096\times 2048$ MIT CCDs,
which were read out in the $2\times 2$ binning mode.

\begin{figure}
\epsfig{figure=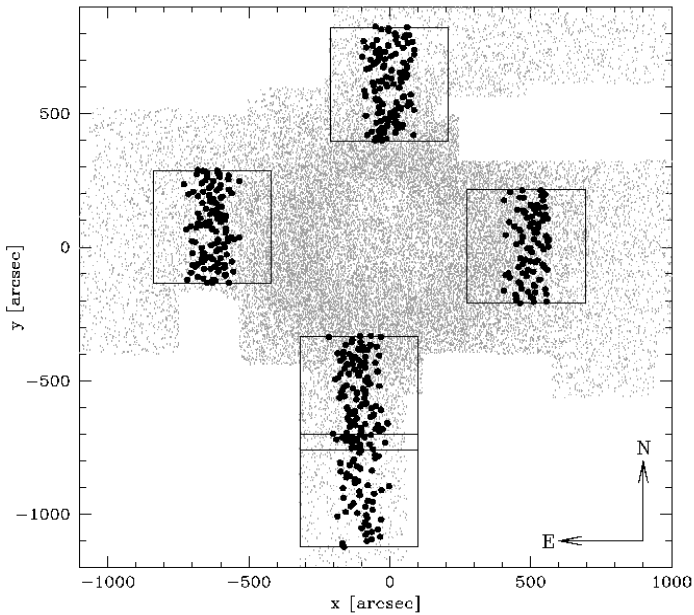,height=8.6cm,width=8.6cm,bbllx=9mm,bblly=10mm,bburx=195mm,bbury=206mm}
\caption{\label{fields} This plot shows in grey the region of \wcen\ that is
covered by Str{\"o}mgren photometry \citep{H/R:00}. All stars with a $V$
magnitude brighter than 19.0 mag and a photometric error less than 0.1 mag have
been plotted. The squared areas indicate the 5 defined fields in which the
candidate stars for the spectroscopy have been selected (bold dots).}
\end{figure}

In a preceding study of \citet{H/R:00}, a large sample of RGB
and MSTO stars in \wcentauri\ was analysed using
Str{\"o}mgren photometry.
The photometry of this study was used to select the candidate stars for the spectroscopy.
Five fields outside the cluster centre were
defined in the region covered by the St{\"o}mgren photometry.
According to their position, the fields were named
`omn' (north), `ome' (east), `oms' (south), `omw' (west) and `omss'
(south-south). Figure~\ref{fields} shows the position of these fields with
respect to the Str{\"o}mgren data.
For this project, we primarily selected stars with magnitudes in the range 17$<V<$18 mag and colours of 0.25$<(b-y)<$0.55 mag,
and those without direct
neighbours to allow a proper sky
subtraction. In total, 11 slit masks have been defined.

\begin{figure}
\psfig{figure=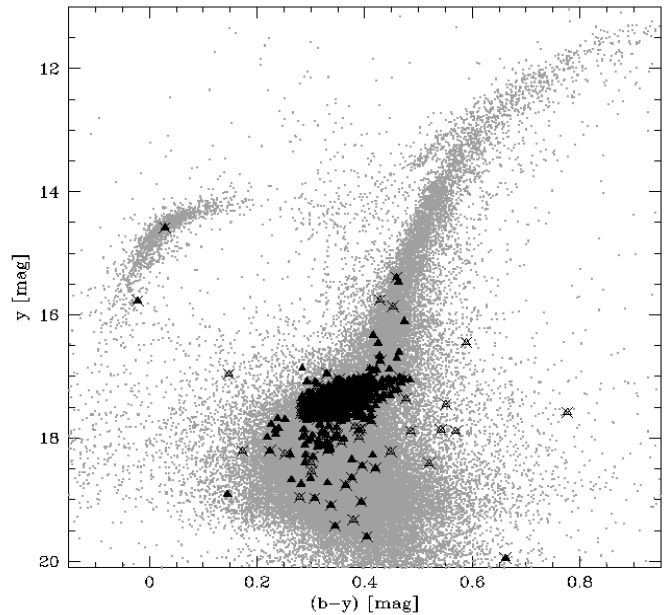,height=8.6cm,width=8.6cm,bbllx=9mm,bblly=10mm,bburx=195mm,bbury=206mm}
\caption{\label{cmd} Position of the programme stars (filled triangles) in the
Str{\"o}mgren CMD of \wcen\ \citep{H/R:00}. The open triangles are those
objects
identified as non cluster members by their radial velocities. All objects
crossed out have been rejected due to the velocity or other selection criteria, such as
quality of the spectrum and close neighbouring stars (see Sect.~\ref{grid}).}
\end{figure}

A slit mask contained 50 to 60 slits with fixed width of 1\arcsec\
and a length of 4-8\arcsec.
Occasionally, one slit covered more than one object.
About 40 stars were observed through more than one slit mask.
Those objects were used to verify the consistency between the observations of
the different nights/masks.

The observations were performed between May 7, 2002 and May 12, 2002. In total,
spectra of $\sim$620 selected objects were observed.
The seeing was between $0\farcs6$ and $1\farcs0$ in all nights.
Each slit mask was observed through two different grisms with the ESO
denotations 1400V+18 and 600I+25 (second order).
In the  $2\times 2$ binning mode the first grism has a dispersion of $0.62\unit{{\AA}\; pix^{-1}}$ in the spectral range 4560 to $5860\unit{\AA}$.
The latter covers the range 3690 to
$4888\unit{\AA}$ with a dispersion of $0.58 \unit{{\AA} \; pix^{-1}}$.
The individual wavelength coverage depends on the position of the slit on the
mask. Lines located in the overlapping wavelength range could be used to check
the consistency between the measurements of the two different grisms.
Each mask was observed two or three times with typical total exposure times of
45 min for the 1400V grism and 60 min for the 600I grism (see Table 1).
Moreover, for each field/mask, calibration exposures such as bias, flat-field,
and lamp spectra were obtained.

During the same observing run, one mask for stars was observed near the MSTO/SGB of the
globular cluster M55.
This chemically homogeneous cluster with a well-known metallicity of
[Fe/H]=$-1.8\unit{dex}$ \citep[e.g.][]{Z:85, C/D:88, R/H/R:99} serves as a
comparison and calibration object.
Furthermore, spectra of 17 standard stars with known abundances
\citep[mostly from][]{C/S/R:01} were obtained with the same grism in the long slit mode.

The data reduction was performed with the IRAF\footnote{IRAF is distributed by 
the 
National Optical Astronomy Observatories, which are operated by the Association
of Universities for Research in Astronomy, Inc., under cooperative agreement 
with the National Science Foundation.}-packages ONEDSPEC and TWODSPEC. 
The mask-exposures were bias corrected, cleaned for cosmic-rays (using BCLEAN
from the STARLINK package), and combined.
The combined spectral images were response-calibrated using dome-flat 
spectra. No absolute flux calibration was performed.
Next, the individual stellar spectra were extracted.
In most cases, object and sky could be extracted from the same slit.
For about 90 stars the sky was taken from another closeby slit and 
subtracted with the SKYTWEAK task.
The lamp exposures were extracted in the same way as the science 
exposures.

For the wavelength calibration, we used 9-15 lines of the elements He, Ne, Hg, and Cd for the 1400V grism. In case of the blue grism (600I, 2nd order), 7-12 He, Hg, and Cd lines were used for the calibration. The overall uncertainty of the wavelength calibration was found to be $0.01\unit{\AA}$ (rms).
For further analyses, all spectra were binned to a spectral scale of 
$1\unit{{\AA} \; pix^{-1}}$ in the ranges 3520 - $5100\unit{\AA}$, for the blue,
and 4340 - $6120\unit{\AA}$ for the red grism.
The final spectral resolution (FWHM) for narrow lines is $\sim 2.5 \unit{\AA}$. 
The signal-to-noise ratio (S/N) varied between 30 and 100
per pixel depending on the grism, wavelength range, and luminosity of the star.
The data for M55 and the standard stars were reduced to 
wavelength-calibrated spectra in the same way.

\begin{figure}
\psfig{figure=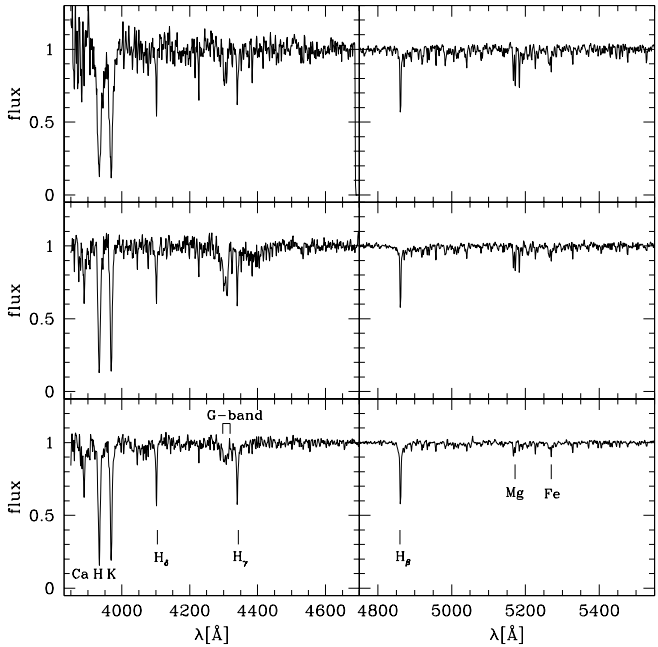,height=8.6cm,width=8.6cm,bbllx=9mm,bblly=60mm,bburx=195mm,bbury=246mm}
\caption{\label{spec} This figure illustrates the different strengths of the 
metal lines in the continuum-normalised spectra of three different stars. 
From top to bottom a metal-rich, an intermediate, and 
a metal-poor star are shown. Note that the three stars have roughly the same 
temperature so that the differences in line strengths are due purely to 
abundance variations.}
\end{figure}


\section{Grid of observed spectra}\label{grid}
The spectra sample included stars with different luminosities and at 
different positions on the slit mask. In order to obtain a more homogeneous 
set, we therefore applied several selection criteria. 

First, we classified the spectra based on their 
overall quality. We sorted out those spectra with errors resulting from 
bad tracing of the spectrum, either due to its position at the border of the 
slit (star not fully in the slit) or due to its faintness. 
Also, spectra of stars that are located closer than 2\arcsec\ to a bright neighbouring star in 
dispersion direction were left out, since it is possible that their spectral 
contributions overlap.
Second, we rejected all but the exposure with the highest S/N of those stars that had been 
observed multiple times through different slit masks. In total, 508 stars (of originally 662 individual spectra) passed these quality checks.

We went on to select the stars based on their observed radial 
velocity. These were determined using the IRAF tasks 
RVIDLINES and FXCOR. Since \wcen\ has a heliocentric velocity of 
$232.3\unit{km\;s^{-1}}$ \citep{H:96}, foreground stars could easily be 
identified by their much lower radial velocities. We selected all those stars with 
observed radial velocities in the range 150 to $350\unit{km\: s^{-1}}$.
This wide range was chosen to account for possible systematic effects in the velocity 
determination that have not been corrected for (e.g. slight mismatches of the arc-lamp and science exposures during day and night observations).
The radial velocity selection resulted in a sample of 483 member stars with
good quality.

\begin{figure}
\psfig{figure=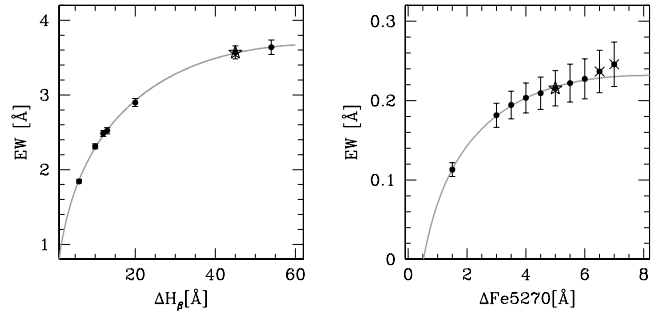,width=8.6cm,bbllx=9mm,bblly=155mm,bburx=195mm,bbury=246mm}
\caption{\label{delta} Effects of the variation in the width of the index
bandpasses on the measured EW for two exemplary lines (H$_\beta$
and Fe5270). The curves are second-order fits to the data points. Excluded data points are crossed out.}
The selected widths of the bandpasses are marked with asterisks.
\end{figure}

The final step was to select only SGB/MSTO-region stars from the member sample.
We included only those stars with magnitudes of $16.8<V_0<18.2\unit{mag}$ and colours of $0.4<(B-V)_0<0.8\unit{mag}$.
Our final sample contains 429 stars.


\section{Measurement of line indices}\label{linemeas}
For the spectroscopic analysis we defined 42 line indices
that include 25 of the strongest iron lines, hydrogen lines, 3 calcium 
and 3 magnesium lines, and the CH-band at $4300 \unit{\AA}$.
The lines were identified and selected by means of the ``Bonner 
Spektralatlas'' \citep{BN-SP-A1, BN-SP-A2} and the online database 
NIST\footnote{NIST (National Institute of Standards and Technology) Atomic 
Spectra Database, Version 2.0 (March 22, 1999), 
\texttt{http://physics.nist.gov/cgi-bin/AtData/main\_asd}}.

For each of these lines a central bandpass region and two flanking continuum 
regions on either side were defined with typical bandwidths of $5-20\unit{\AA}$.
We chose the widths of the central bands in such a way that only the line to be 
measured was included.
This approach was problematic for the strong Balmer and calcium lines, though.
These lines have extended wings in which one can find lines of other elements.
In case of the weaker Fe lines it was very important to minimise the effects of
neighbouring lines, since even weak neighbouring lines can have a considerable 
influence on the measured equivalent width (EW).
Figure~\ref{delta} illustrates the effect of varying the width of the 
central line bandpass on the measured line indices using the example of the ${\rm H_{\beta}}$ and the Fe5270 line. With larger width of the central line bandpass ($\Delta \rm H_{\beta}$ and $\Delta \rm Fe5270$), the EW first increases very rapidly and only marginally after the whole line is covered by the bandpass. With the further increase of the bandwidth the influence of included neighbouring lines can be seen, especially for the weaker Fe5270 line.
We selected the bandwidth large enough to cover the whole line but as well small enough to prevent major effects by neighbouring lines.
In the case of very close lines of the same element the central bandpass was 
extended to cover all of them. These indices were labelled with ``B\_" (=band).

\begin{figure}
\psfig{figure=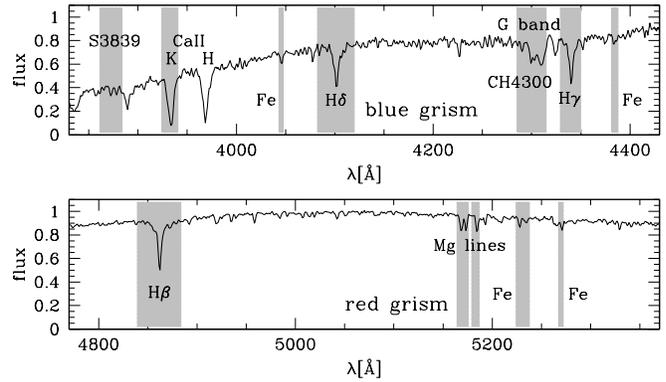,height=4.7cm,width=8.6cm,bbllx=9mm,bblly=145mm,bburx=195mm,bbury=246mm}
\vspace{0.4cm}
\caption{\label{bands} The blue and red part of a spectrum of a typical SGB 
star of $\omega$ Cen is shown. The maximum count rate was normalised to unity. 
The shaded regions mark most of the bandpasses that were used to measure iron 
and Balmer line indices as well as CN, CH and $\alpha$ indices.}
\end{figure}

Following the specifications in the Lick system, the EW of an 
index was determined by 
the flux difference between the pseudo-continuum and the spectrum in the range 
of the bandpass \citep[e.g.][]{W/F:94}.
The pseudo-continuum is defined as a line drawn between the midpoints of the 
two continuum bandpasses.
There were 6 line indices defined in the overlapping wavelength region of the 
two grisms.  Those indices were measured in both the red and blue 
spectra individually, and the indices are correspondingly labelled `r' and `b' .
Figure~\ref{bands} shows the location of some of the defined 
line indices, and Tables 2 and 3 summarise all indices.

For the automatic measurement of the line indices we used a program 
(\texttt{gonzo.prl}) provided by Thomas H. Puzia\footnote{\texttt{tpuzia@stsci.edu}}.
This algorithm measures line indices of abundance features on ASCII files for 
any kind of spectra.
The overall errors are estimated based on the Poisson noise of the spectra, on the errors in the velocity determination, and by the uncertainties of the sky spectrum.

\begin{table}[ht!]
\caption{\label{tab2}Definition for line indices in the blue grism}
\centering
\begin{minipage}[t]{8cm}
\hspace*{-0.4cm}
{\tiny
\begin{tabular}[l]{crlrlrl}
\hline\hline
 Index & \multicolumn{2}{c}{central bandpass} & 
\multicolumn{2}{c}{blue continuum} & \multicolumn{2}{c}{red continuum} \\
  & l. limit & u. limit & l. limit & u. limit & l. limit & u. limit  \\
 \hline
  CN\_Fe & 3853.5 & 3875.0 & 3847.0 & 3853.0 & 3909.0 & 3923.0 \\ 
CaII\_K & 3924.0 & 3942.0 & 3909.0 & 3917.0 & 3988.0 & 3995.0  \\ 
CaII\_H & 3954.0 & 3987.0 & 3909.0 & 3917.0 & 3988.0 & 3995.0  \\ 
Mn4032 & 4027.5 & 4036.5 & 4018.5 & 4027.0 & 4037.0 & 4042.5  \\ 
Fe4045 & 4043.0 & 4048.0 & 4037.0 & 4042.0 & 4049.0 & 4061.0  \\ 
Fe\_B01 & 4061.5 & 4080.0 & 4049.0 & 4061.0 & 4115.5 & 4128.5  \\ 
H$_\delta$ & 4082.0 & 4120.0 & 4049.0 & 4061.0 & 4147.0 & 4166.0  \\ 
Fe4132 & 4129.0 & 4136.5 & 4116.0 & 4128.5 & 4137.0 & 4141.0  \\ 
Fe4143 & 4141.5 & 4146.0 & 4137.0 & 4141.0 & 4146.5 & 4166.0  \\ 
Fe\_B02 & 4184.0 & 4206.0 & 4146.5 & 4166.0 & 4206.5 & 4213.5  \\ 
Fe4215 & 4213.5 & 4218.0 & 4206.5 & 4213.5 & 4230.5 & 4245.0  \\ 
 Fe\_Ca & 4223.5 & 4229.5 & 4206.5 & 4213.5 & 4230.5 & 4245.0  \\ 
 G4300 & 4282.0 & 4318.0 & 4239.0 & 4269.0 & 4357.0 & 4370.0  \\ 
    CH & 4322.0 & 4328.0 & 4239.0 & 4269.0 & 4357.0 & 4370.0  \\ 
H$_\gamma$ & 4329.0 & 4350.0 & 4239.0 & 4269.0 & 4364.0 & 4371.0  \\ 
Fe4376 & 4373.0 & 4380.0 & 4357.0 & 4370.0 & 4409.0 & 4413.0  \\ 
Fe4383 & 4381.0 & 4388.0 & 4355.0 & 4370.0 & 4409.0 & 4413.0  \\ 
Ti4395 & 4391.5 & 4397.0 & 4355.0 & 4370.0 & 4409.0 & 4413.0  \\ 
Fe4405 & 4402.5 & 4409.0 & 4355.0 & 4370.0 & 4409.0 & 4413.0  \\ 
Ca4455 & 4452.0 & 4457.0 & 4432.0 & 4451.0 & 4458.0 & 4473.0  \\ 
bTi4534 & 4530.0 & 4536.5 & 4504.0 & 4524.0 & 4537.0 & 4545.0  \\ 
bFe4549 & 4546.0 & 4552.0 & 4537.0 & 4545.0 & 4552.0 & 4570.0  \\ 
bH$_\beta$ & 4839.0 & 4884.0 & 4815.0 & 4829.0 & 4895.0 & 4905.0  \\ 
bFe4920 & 4915.5 & 4922.0 & 4901.5 & 4915.0 & 4941.5 & 4952.0  \\ 
bFe4957 & 4954.0 & 4960.0 & 4942.0 & 4952.0 & 4961.0 & 4980.0  \\ 
bFe\_B03 & 5003.5 & 5021.0 & 4992.0 & 5003.0 & 5021.0 & 5038.5  \\
\hline\hline
\end{tabular}}
\end{minipage}
\end{table}

\begin{table}[h!]
\caption{\label{tab3}Definition for line indices in the red grism}
\centering
\begin{minipage}[t]{8cm}
\hspace*{-0.4cm}
{\tiny
\begin{tabular}[l]{crlrlrl}
\hline\hline
 Index & \multicolumn{2}{c}{central bandpass} & 
\multicolumn{2}{c}{blue continuum} & \multicolumn{2}{c}{red continuum} \\
  & l. limit & u. limit & l. limit & u. limit & l. limit & u. limit  \\
 \hline
rTi4534  & 4530.0 & 4536.5 & 4504.0 & 4524.0 & 4537.0 & 4545.0  \\ 
rFe4549  & 4546.0 & 4552.0 & 4537.0 & 4545.0 & 4552.0 & 4570.0  \\ 
rH$_\beta$  & 4839.0 & 4884.0 & 4815.0 & 4829.0 & 4895.0 & 4905.0  \\ 
rFe4920  & 4915.5 & 4922.0 & 4901.5 & 4915.0 & 4941.5 & 4952.0  \\ 
rFe4957  & 4954.0 & 4960.0 & 4942.0 & 4952.0 & 4961.0 & 4980.0  \\ 
rFe\_B03  & 5003.5 & 5021.0 & 4992.0 & 5003.0 & 5021.0 & 5038.5  \\ 
Fe5041  & 5039.0 & 5044.0 & 5021.0 & 5039.0 & 5054.0 & 5064.0  \\ 
Fe5139  & 5136.5 & 5141.5 & 5116.0 & 5136.0 & 5144.0 & 5161.0  \\ 
Mg\_1+2  & 5164.0 & 5177.0 & 5144.0 & 5161.0 & 5210.0 & 5223.5  \\ 
Mg5183  & 5179.0 & 5187.0 & 5144.0 & 5161.0 & 5210.0 & 5223.5  \\ 
Fe\_B04  & 5224.0 & 5238.0 & 5210.0 & 5223.5 & 5239.0 & 5259.0  \\ 
Fe5270  & 5267.0 & 5272.5 & 5240.0 & 5259.0 & 5286.0 & 5311.0  \\ 
Fe5328  & 5322.5 & 5331.0 & 5299.5 & 5316.0 & 5351.0 & 5365.5  \\ 
Fe5371  & 5368.0 & 5373.5 & 5351.0 & 5365.5 & 5375.0 & 5394.0  \\ 
Fe5397  & 5394.0 & 5400.0 & 5375.0 & 5394.0 & 5410.0 & 5422.0  \\ 
Fe5405  & 5402.0 & 5408.0 & 5375.0 & 5394.0 & 5410.0 & 5422.0  \\ 
Fe\_B05  & 5426.0 & 5437.0 & 5410.0 & 5422.0 & 5459.0 & 5471.0  \\ 
Fe5446  & 5442.0 & 5450.0 & 5410.0 & 5422.0 & 5459.0 & 5471.0  \\ 
Fe5455  & 5453.0 & 5458.0 & 5410.0 & 5422.0 & 5459.0 & 5471.0  \\ 
Mg5528  & 5523.0 & 5533.0 & 5512.0 & 5522.0 & 5535.0 & 5546.0  \\ 
 Na\_D1  & 5888.0 & 5893.0 & 5863.0 & 5877.0 & 5900.0 & 5910.0  \\ 
 Na\_D2  & 5894.0 & 5898.5 & 5863.0 & 5877.0 & 5900.0 & 5910.0  \\
\hline\hline
\end{tabular}}
\end{minipage}
\end{table}

\begin{figure}
\psfig{figure=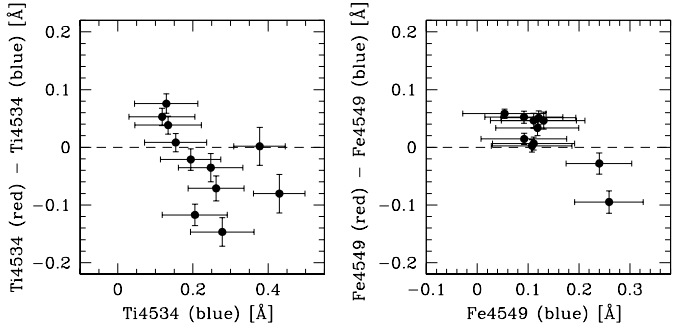,width=8.6cm,bbllx=9mm,bblly=155mm,bburx=195mm,bbury=246mm}
\caption{\label{RB} Comparison of indices measured in both grisms (Ti4534 and
Fe4549).} 
\end{figure}

We performed tests to check the self-consistency of the measurements and to estimate the errors.
As mentioned in Sect.~\ref{obsred} the covered spectral ranges of the two 
grisms slightly overlap. In Fig.~\ref{RB} we compare the measured equivalent 
widths for 2 of the 6 lines (Ti4534 and Fe4549) located in this region.
Although the scatter for the Ti4634 is quite large and there might be
some overestimation of the line strength of Fe4549 in the red spectrum,
we consider the agreement acceptable, in particular when taking into
account that the lines are very weak and located close to the edge of
the spectra.
The tracing of the spectrum in these regions was difficult during the data reduction.
Bad tracing leads to a misdefinition of the pseudo continuum and therefore to
mismeasurements of the line strengths.
     
\begin{figure}
\psfig{figure=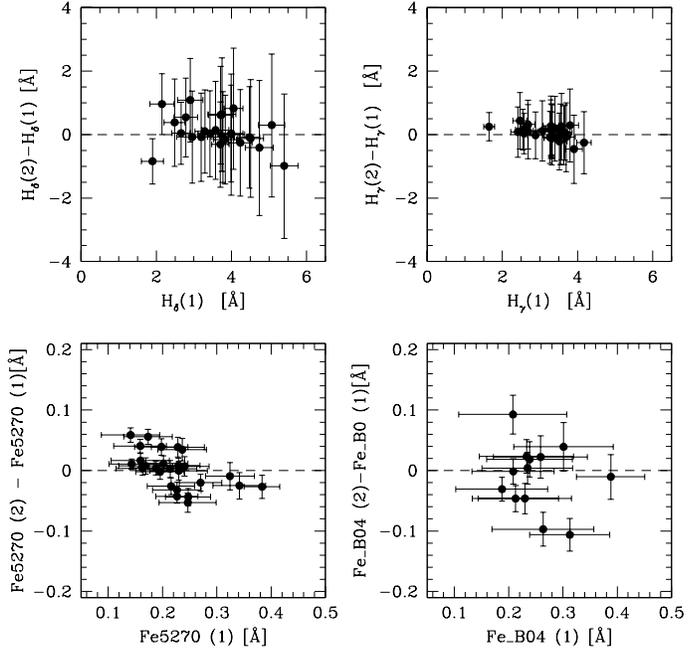,width=8.6cm,bbllx=9mm,bblly=60mm,bburx=195mm,bbury=246mm}
\caption{\label{dop} Comparison of indices (H$_\delta$,  H$_\gamma$, Fe5270, and
Fe\_B04) measured in different spectra of the same star. The 1:1 correlation is
indicated by the dashed line. All line indices affirm the very good consistency
between the different nights/masks.}
\end{figure}

Another check on the accuracy of the measurements is provided by stars that
were observed through more than one slit-mask.
For those objects the defined line-indices were measured twice.
Accordingly, one would expect identical values for the EW.
As an example, Fig.~\ref{dop} shows the comparison of the indices H$_\delta$,  
H$_\gamma$, Fe5270, and Fe\_B04 that were measured in different spectra of
identical stars. In these graphs only data with a fractional error of less than 
50\% of the index value are shown. 
For all indices one sees good consistency within the errors.

In order to calculate a more accurate indicator for hydrogen and 
iron, we averaged several individual line indices to a single-line index for a given element.
For hydrogen, we combined 2 of the 3 prominent Balmer lines:
\begin{equation}
\rm{\langle H \rangle = (H_{\gamma}+H_{\delta})/2}.
\end{equation}
During the analysis we found that index ${\rm H_{\beta}}$ showed strong discrepancies as compared to ${\rm H_{\gamma}}$ and ${\rm H_{\delta}}$. ${\rm H_{\beta}}$ is the strongest 
of these three Balmer lines and has very extended wings at these temperatures 
and surface gravities. It is very likely that the influence of lines in those 
wings and non-linearities effects cause the observed deviance.
We therefore excluded ${\rm H_{\beta}}$ from the combined H-index.

In order to get an average iron index, 6 of the strongest iron lines and
one magnesium line of similar strength were combined. Before averaging
the 7 indices, they were weighted in such a way that their line
strengths are of the same order. In other words, the weaker
indices were multiplied by larger weighting factors than the stronger
indices. The weighting factors were defined in the following way:
the line strength versus $\langle{\rm H}\rangle$ distribution of each
individual index was fitted by a linear least-square fit, i.e. Fe4045 $=
a\cdot\langle{\rm H}\rangle+b$. The reciprocal of the fit value at \mh\ =
3 defines the weighting factor $w$, i.e. $w_{\rm Fe4045} = 1/(3a+b)$.
With this definition, the combined iron index scatters around 1 at \mh\ =
3 (see e.g. Fig.~\ref{calib}). 
The formula for the combined iron index
showing the weighting factors of the individual indices is:
\begin{eqnarray}
\nonumber
\rm{\langle Fe \rangle} &=& \rm{(4.13 \cdot Fe4045+2.91 \cdot 
Fe\_Ca+3.67 \cdot
Fe4383} \\
\nonumber
& & \rm{+3.42 \cdot Mg5183+4.08 \cdot Fe\_B04+4.04 \cdot Fe5270} \\
& & \rm{+4.67 \cdot Fe5328)/7}.
\end{eqnarray}
This iron index was used in combination with the combined hydrogen index (see above) for the subsequent analysis.

\section{Analysis of the line indices}\label{quali}
\subsection{Luminosity, surface gravity, and temperature effects}\label{effects}
Besides the element abundance, the strength of the iron lines depends on 
other stellar parameters. The most important ones here were the effects 
of effective temperature ($T_{\rm eff}$) (or correspondingly the colour of the 
star) and the surface gravity (\logg ) of the star.

In order to analyse the impact of these parameters on the line strengths, we used
synthetic spectra provided by \citet{BJ:00}. We adjusted their step size 
($0.5\unit{{\AA} \; pix^{-1}}$) and resolution ($\sim 1\unit{\AA}$) to the observed data 
($1\unit{{\AA} \; pix^{-1}}$ and $\sim 2-2.5\unit{\AA}$).
Additionally, we scaled the synthetic counts to the order of magnitude of the 
observational data. We used spectra of different metallicities
($-2.0, -1.5,-1.0$ and $-0.5\unit{dex}$), effective temperatures ($5000 - 
7000\unit{K}$), and surface gravities (range $3.2$ to $4.6$ dex). On the 
synthetic spectra, we then measured the line indices introduced above in the same 
way as for the stellar spectra of \wcen\ .

\begin{figure}
\psfig{figure=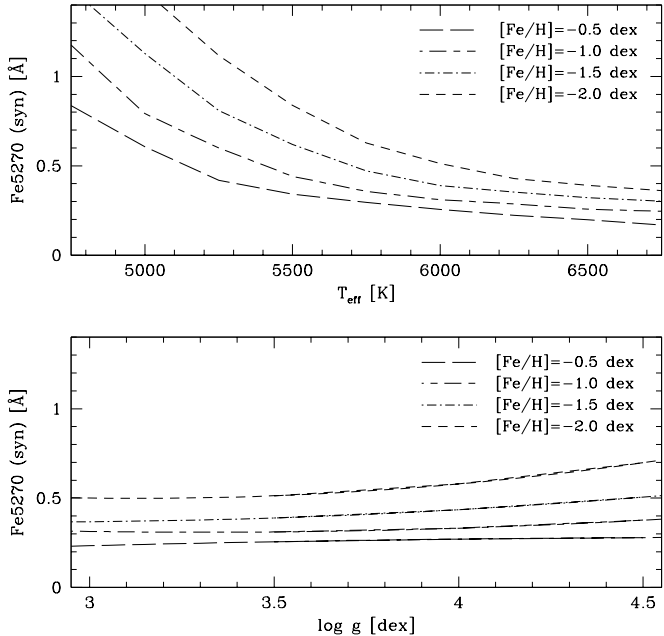,height=8.6cm,width=8.6cm,bbllx=9mm,bblly=60mm,bburx=195mm,bbury=246mm}
\caption{\label{eff} Effects of temperature and surface gravity. The upper 
panel shows the measured Fe5270 index as a function of the effective 
temperature for four different iron abundances ([Fe/H]), with constant $\logg\ 
=3.5$. Due to the temperature effect, the measured EW increases with 
decreasing effective temperature for constant [Fe/H]. In the lower panel the 
Fe5270 index is plotted versus \logg\ in case of four different [Fe/H] and a 
constant temperature $\Teff\ =6000\unit{K}$.}
\end{figure}

\begin{figure}
\psfig{figure=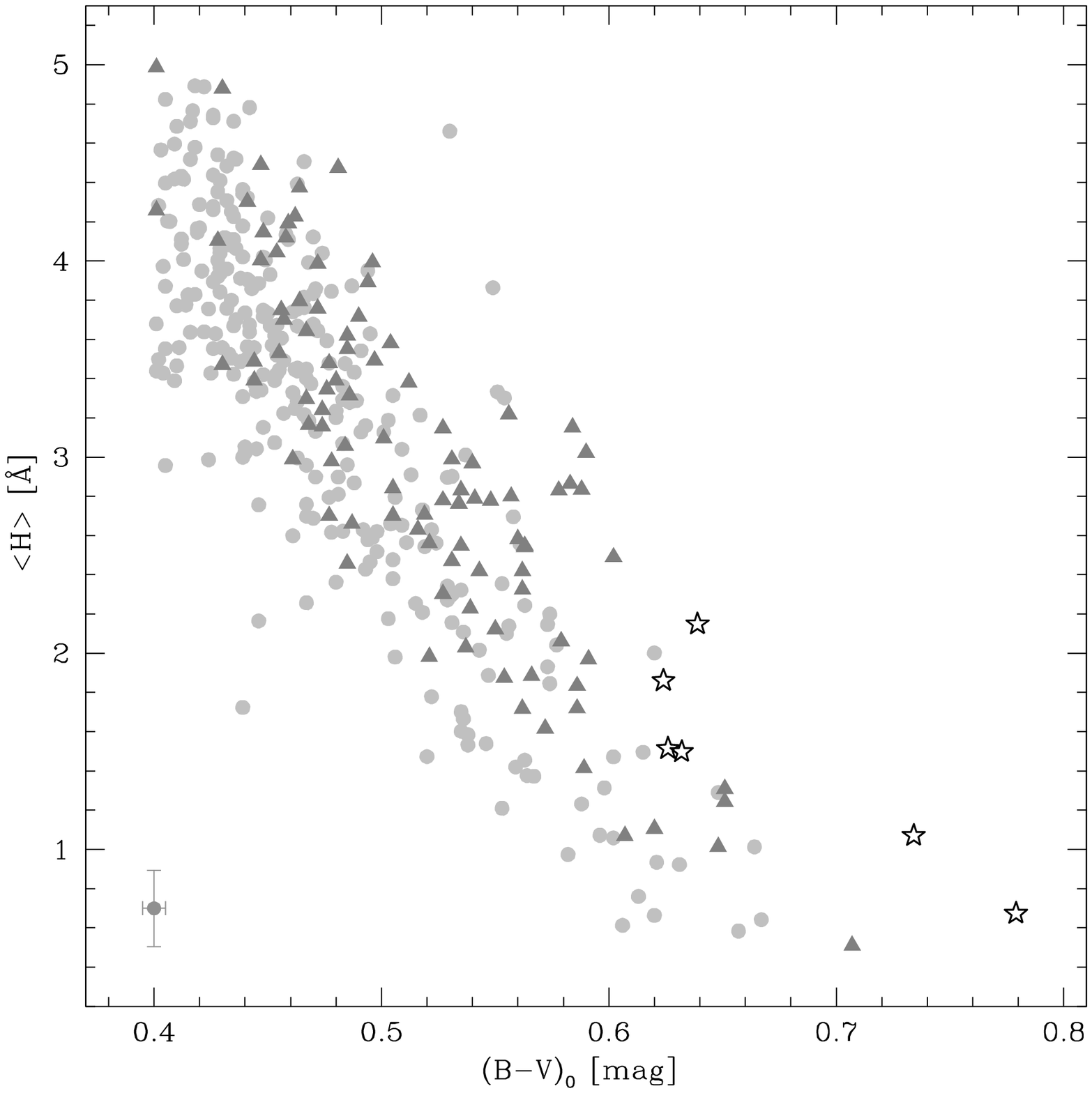,height=8.6cm,width=8.6cm,bbllx=9mm,bblly=60mm,bburx=195mm,bbury=246mm}
\caption{\label{H_BV} This figure shows the coherency between mean Balmer index
\mh\ and the $(B-V)_0$ colour for the \wcen\ data. \mh\ mainly depends on 
temperature, as demonstrated by the relatively small scatter in this diagram, and
therefore can be used as a temperature indicator. The  scatter in this
diagram is due to small metallicity effects on \mh\ and $(B-V)$:
the light grey dots indicate stars with iron abundances between $-2.0$ and 
$-1.4\unit{dex}$; the grey triangles are for those with 
$-1.4<$[Fe/H]$<-1.0\unit{dex}$, and the dark asterisks are for stars with
$-1.0<$[Fe/H]$<-0.5\unit{dex}$.}
\end{figure}

As an example, the dependence of the strongest iron line Fe5270 on \Teff\ and 
\logg\ is shown in Fig.~\ref{eff}. One can see that the temperature is a major 
influence on the line index at those parameter values ($3.2<$ \logg\ $< 4.1$ 
dex and $5400<$ \Teff\ $<6000$ K).
To correct for the effects of temperature, we used the combined Balmer index \mh ,
which shows a strong correlation with the $(B-V)_0$ colour (see 
Fig.~\ref{H_BV}). The scatter in the \mh\ -$(B-V)_0$- diagram is mainly caused 
by an additional, but weak, dependence on metallicity and \logg . 

\subsection{Comparison with M55 spectra}\label{m55spec}
\begin{figure}
\psfig{figure=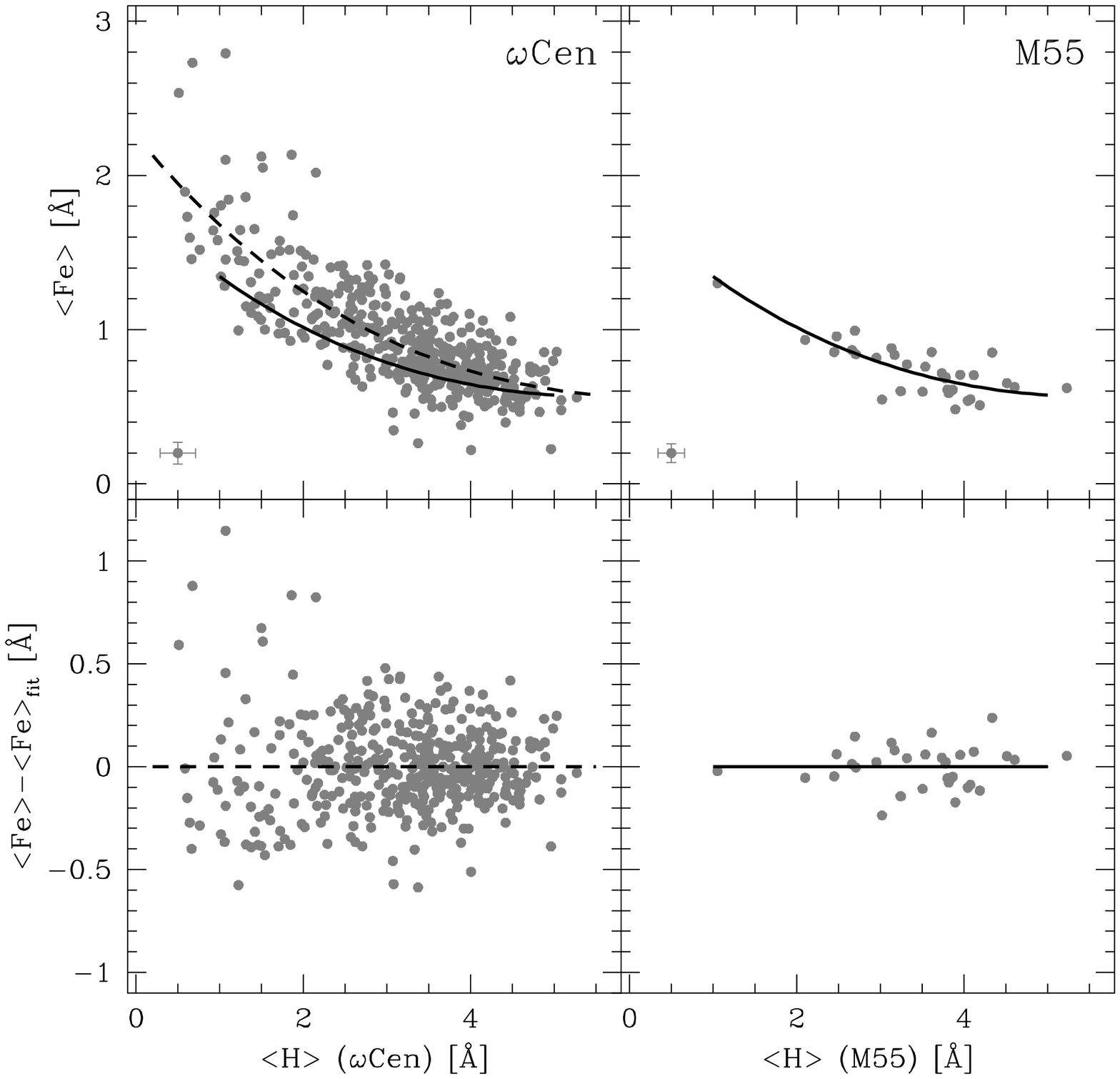,height=8.6cm,width=8.6cm,bbllx=9mm,bblly=60mm,bburx=195mm,bbury=246mm}
\caption{\label{M55} The upper two panels show the \mfe- \mh- diagrams of 410
MSTO/SGB stars in \wcen\ (left) and 36 M55 stars (right). The \wcen\
data points show a much larger scatter in the \mfe\ index. The solid line is an
exponential fit to the M55 data points and therefore marks the position of
stars with [Fe/H]$=-1.8 \unit{dex}$ in both diagrams. The dashed line in the
left panel is an exponential fit to the \wcen\ data points. In the lower left
corner the average error is shown. The two lower figures show the scatter of the
\wcen\ and M55 data points around the fit curves. The scatter in \wcen\ is much
larger than in M55 and can only be explained by the wide intrinsic
metallicity spread.}
\end{figure}

One possibility for deriving absolute metallicities is comparing the line strengths of the stars in \wcen\ with those in M55. This globular cluster has a similar 
distance to the Sun ($d_{M55}=5.4\unit{kpc}$ and $d_{ \omega Cen }=5.2 
\unit{kpc}$; \citealt{H:96}); i.e. the MSTO stars
have luminosities comparable to those of \wcen\  ($V_{\rm MSTO} \sim 17.5 
\unit{mag}$), and thus their spectra are of the same quality in terms of S/N.
After clearing the M55 sample of non-members (by radial velocities) and 
selecting 
stars with $0.35<(B-V)_0<0.68\unit{mag}$ and $17.1<V_0<17.8\unit{mag}$ we 
measured the defined line indices on the remaining 36 spectra.
As for \wcen ,  we calculated a combined Fe and H index.
The upper panels of Fig.~\ref{M55} show the \mfe -\mh -diagrams for both \wcen\ and M55.
As the M55 stars have a well-known iron abundance of [Fe/H]$=-1.8\unit{dex}$, 
their average position in the \mfe -\mh -diagram can serve as reference in 
the \wcen\ plot.

Figure~\ref{M55} shows that \wcen\ has a much larger variation in iron 
abundances than M55. Furthermore, one can see that most of the stars have
higher abundances than [Fe/H]$=-1.8\unit{dex}$. This result is in qualitative 
agreement with former results for RGB stars \citep[e.g.][]{N/DC:95, S/K:96}.

Although the \mfe -\mh -relation of M55 stars cannot be used for an absolute 
calibration of the most metal-rich stars in \wcen , it is very useful to 
estimate the accuracy of the index measurements.
The analysis of the scatter of the M55 stars around the fit (upper right panel) reveals a standard deviation of $\sigma _{M55} 
({\rm Fe})\sim 0.10\unit{\AA}$ (see Fig.~\ref{M55} lower panel).
The EW is proportional to the number of absorbing atoms (for non-saturated lines, which can be assumed for the weak Fe lines), i.e. EW$_{\rm Fe} \sim N_{\rm Fe}$ and [Fe/H]$\sim  {\rm log}N_{\rm Fe}$.
By defining an upper and lower error in EW 
((${\rm EW}+\Delta{\rm EW}$) and (${\rm EW}-\Delta{\rm EW}$)) we estimated the internal
error in [Fe/H] to be $0.12\unit{dex}$.

In the lower left panel in Fig.~\ref{M55} we plotted the scatter around the 
fit curve of the \wcen\ data (upper panel of Fig.~\ref{M55}).
The standard deviation for this fit is $\sigma _{\omega \; Cen} 
({\rm Fe})\sim 0.22\unit{\AA}$, significantly larger than that for M55. 
This a clear indication of an intrinsic broad scatter in [Fe/H] in the region 
of the MSTO and SGB.

\subsection{Comparison with standard stars}\label{stdstars}
In addition to the observations of M55, we obtained spectra of stars with known 
stellar parameters. These are referred to as `standard' stars in the following. 
The spectra of these stars have higher S/N values than the globular cluster 
spectra (typically $\sim 80-150$ at $4500\unit{\AA}$ and $\sim 180-220$ at 
$5100\unit{\AA}$).
From the sample of 17 observed standard stars, we selected 11 stars with
spectral types of SGB/MSTO region stars.
Table~\ref{tabstd} summarises their properties.
On these spectra we again measured the line indices and calculated the average 
iron and hydrogen indicators \mfe\ and \mh\ as defined in 
Sect.~\ref{linemeas}.

For more clarity, we divided the standard stars into four groups
according to the [Fe/H] values given by the literature.
All stars with Fe abundances higher than $-0.95\unit{dex}$ are classified as
most metal-rich, followed by stars with metallicities in the range
$-1.3<$[Fe/H]$<-0.95\unit{dex}$ and $-1.8<$[Fe/H]$<-1.3\unit{dex}$.
Stars with [Fe/H]$<-1.8\unit{dex}$ are termed as `metal-poor'.

In Fig.~\ref{calib}, we plot the \mfe -\mh\ diagram for \wcen , together with
the data points of the standard stars. The positions of the different groups
of standard stars mark the relation between \mh, \mfe, and [Fe/H].
The metal-poor standards form a lower limit for the distribution of \wcen\
stars, whereas the most metal-rich standards form an upper limit.
Most stars of \wcen\ are associated with an Fe abundance between $-2.0$ and
$-1.3\unit{dex}$.

\subsection{Comparison with synthetic spectra}\label{synth}
Another independent calibration method is the use of synthetic spectra (see
Sect.~\ref{effects}). By measuring the same line indices on these spectra and by 
calculating \mfe\ and \mh\ we were able to derive synthetic iso-metallicity
curves in the \mfe -\mh -diagram. Only synthetic spectra with \Teff\ and
\logg\ values that are consistent with MSTO/SGB stars (adopted from the
isochrone set by \citet{K/D/Y/A:02}) were considered. In this way,
the iso-metallicity curves are implicitly corrected for the \logg\ effect
of MSTO/SGB stars (see Fig.~\ref{eff}).

\begin{table}
\caption{\label{tabstd} General properties of the standard stars used for the
metallicity calibration. The values are taken from the SIMBAD database
(\texttt{http://simbad.u-strasbg.fr/Simbad})}
\centering
\begin{minipage}[t]{8cm}
\hspace*{-5mm}
\begin{tabular}{crccrrc}
\hline\hline
Name & $V$ &  $B-V$ & Type & \logg & [Fe/H] & H$_\beta$ \\
 & [mag] & [mag] &  & [dex] & [dex] & [mag] \\
 \hline
BD-032525 &  9.67 & 0.48 & F3 & 3.60 & $-$1.90 & 2.59 \\
BD-052678 & 11.92 & 0.59 & F7 & 4.43 & $-$1.89 & 2.59 \\
BD-084501 & 10.59 & 0.59 & F8 & 4.00 & $-$1.40 & 2.59 \\
BD-133442 & 10.37 & 0.32 & F  & 4.20 & $-$2.72 & -- \\
HD089499 &  8.68 & 0.74 & F8 & 3.00 & $-$2.23 & 2.54 \\
HD097916 &  9.17 & 0.38 & F5V & 4.20 & $-$0.92 & 2.64 \\
HD179626 &  9.14 & 0.53 & F7V & 3.70 & $-$1.26 & 2.59 \\
HD192718 &  8.41 & 0.54 & F8 & 4.00 & $-$0.74 & -- \\
HD193901 &  8.67 & 0.53 & F7V & 4.58 & $-$1.03 & 2.57 \\
HD196892 &  8.23 & 0.50 & F6V & 4.14 & $-$1.00 &  2.59   \\
HD205650 &  9.00 &  0.59 & F8V & 4.38 & $-$1.03 & -- \\
\hline\hline
\end{tabular}
\end{minipage}
\end{table}


\section{Absolute calibration of the abundances}\label{abscal}
\begin{figure}
\psfig{figure=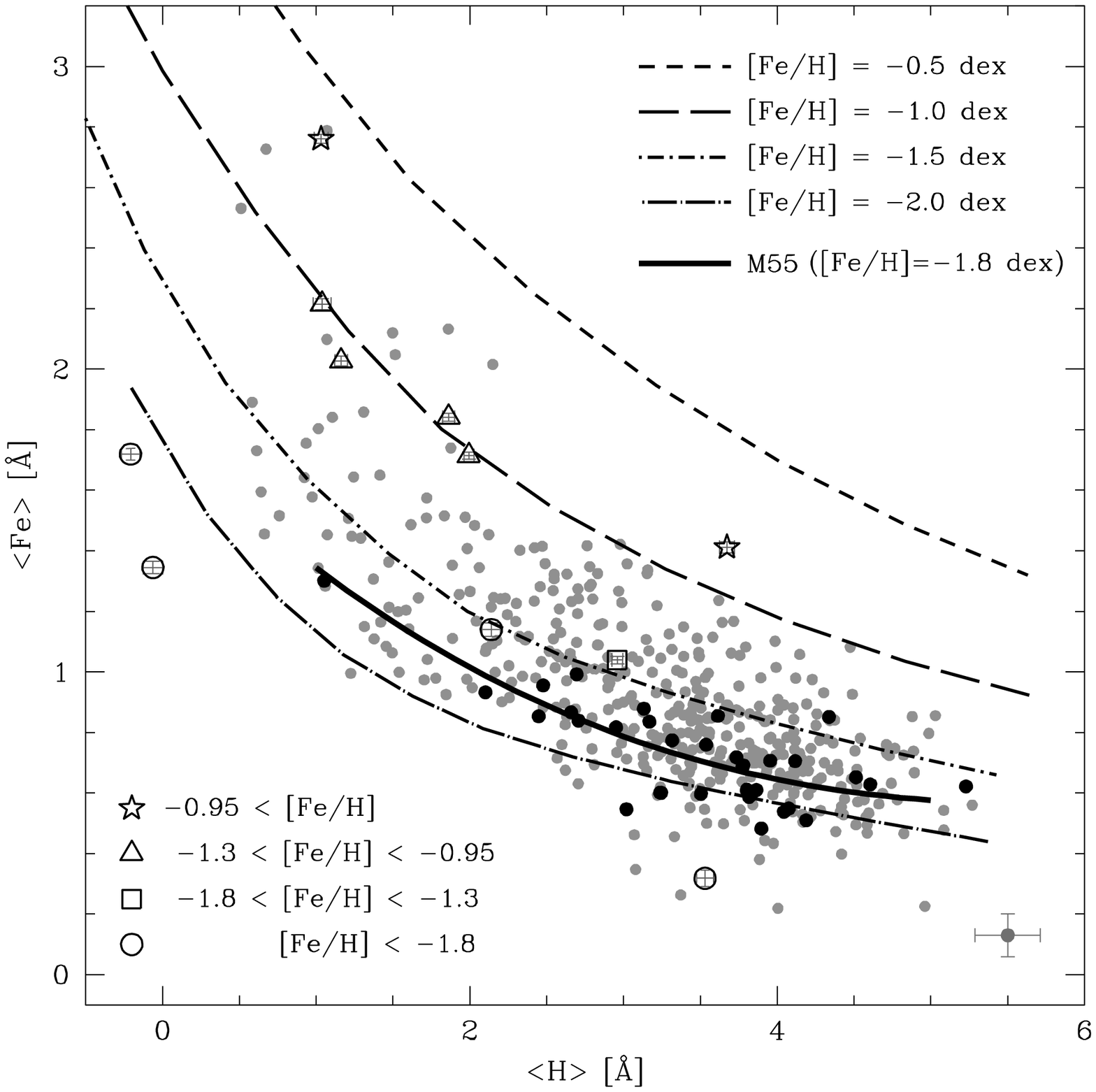,height=8.6cm,width=8.6cm,bbllx=9mm,bblly=60mm,bburx=195mm,bbury=246mm}
\caption{\label{calib} This plot shows the distribution of 410 MSTO/SGB stars 
in \wcen\ (grey dots) in the \mfe -\mh -diagram with respect to all three 
independent calibrators. The positions of the 11 standard stars are plotted for
four metallicity ranges as indicated in the lower left. The bold curve is the 
fit of the M55 datapoints as derived in Sect.~\ref{m55spec}. The dashed lines
are the iso-metallicity curves as calculated from the synthetic spectra. These 
lines are constructed in such a way that they account for the \logg\ effect on 
the SGB where \logg\ decreases steadily from 4.1 to 3.5 with increasing \mh. 
All independent calibration methods are consistent with each other and allow 
for an assignment of absolute iron abundances. The average error bar is given in 
the lower right.}
\end{figure}

The absolute calibration of iron abundances was done by using all calibrators introduced above. The distribution of the standard stars, the M55 fit 
curve, and the 
iso-metallicity curves of the synthetic spectra are shown in Fig.~\ref{calib}.
The metallicity positions of all calibrators are consistent with each other,
the \wcen\ data points are plotted for reference.

In order to derive the absolute [Fe/H] abundances for the \wcen\ stars from 
their \mh\ and \mfe , we established an analytical relation of the form 
[Fe/H]=f(\mfe , \mh ). A polynomial fit of 4th order in both coordinates was used with 
full cross-terms. 
The determined metallicities were analysed in our previous paper \citep{H/K:04}.
As our sample is strongly biased by the selection of the target stars, our metallicity distribution
is not representative for \wcen ; hence we do not show the biased
metallicity distribution here.

The total errors in [Fe/H] were derived from the errors in the measurements of the individual
line indices and the error of the derived relation between \mh , \mfe\ and [Fe/H].
The former are based on the Poisson noise of the spectra, the errors in the velocity determination and on the uncertainties on the sky spectrum, whereas the latter include the uncertainties of the polynomial fit.

\begin{figure}
\psfig{figure=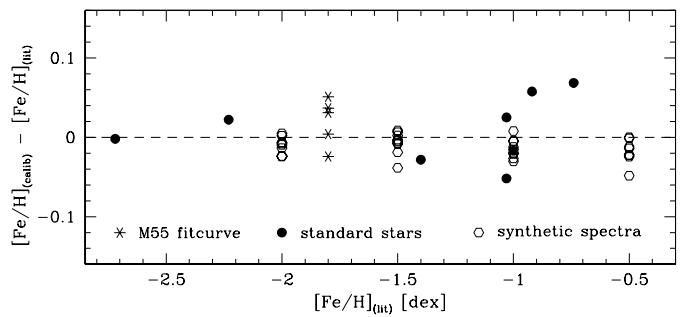,width=8.6cm,width=8.6cm,bbllx=9mm,bblly=155mm,bburx=195mm,bbury=246mm}
\caption{\label{fit} Scatter of the calibrated iron abundances versus their
literature values. The standard stars are marked as bold dots, the synthetic
data points as open symbols. Five sample points out of the M55 fit curve
are given as asterisks.}
\end{figure}

In order to estimate the error resulting from the calibration method, we
compared the [Fe/H] literature values with those calculated by our established
analytical relation for the standard stars, the M55 data, and the synthetic
data points. Figure~\ref{fit} shows the difference between the [Fe/H]
values as derived from our calibration
versus the literature values. For the case of M55, we plot five
values with different \mh . The scatter of the calibrators
around the calibration relation is $\sigma _{\rm calib}({\rm Fe})\sim
0.04\unit{\AA}$.

\begin{figure}
\psfig{figure=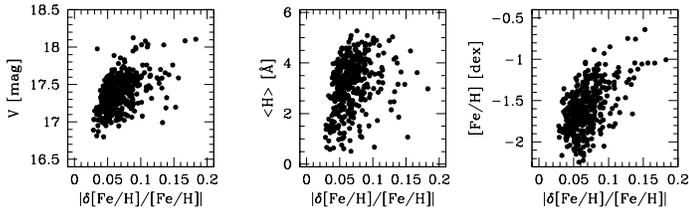,width=8.6cm,width=8.6cm,bbllx=15mm,bblly=60mm,bburx=195mm,bbury=121mm}
\caption{\label{err} The dependency of relative error in [Fe/H] on V, \mh\ (i.g. \Teff ), and [Fe/H]. For brighter and cooler stars the determined [Fe/H] values have the smallest relative errors.}
\end{figure}

The total errors in [Fe/H], $\delta$[Fe/H], were found to be of the order of $0.1 - 0.3 \unit{dex}$.
As the most luminous stars ($\approx 16.5 \unit{mag}$) in our sample provide the spectra with the highest S/N ratios, the measured indices on those spectra have the smallest errors. The mean relative error in [Fe/H] for stars between V=16.6 and 17.0 mag is $\delta$[Fe/H]/[Fe/H]$=5.1\%$, whereas for fainter stars ($18.0 <V < 18.4$) this error is $12.5\%$.
That temperature has a major effect on the strength of the Fe lines (Sect.~\ref{effects}) is reflected in the relative errors of [Fe/H].
Figure~\ref{err} shows the relation of the relative error in [Fe/H] with \mh\ i.e. temperature. For lower \mh\ (lower \Teff), the relative errors are smallest. 
In other words, for a given [Fe/H] the observed iron lines are stronger the lower the effective temperature \Teff\ (in the temperature regime considered by us).
Consequently, the total errors in [Fe/H]
are mainly influenced by the luminosity and temperature of the individual
stars (see Fig.~13).

\section{Relative abundances of light elements in the different sub-populations}\label{other}
In addition to Fe, we also measured indices of the Ca and Mg lines and of the CN and CH
bands. The calcium index \mca\ is just the strength of the Ca K line. The
magnesium index \mmg\ is the average of the Mg indices Mg\_1+2 and Mg5183
(see Table 3).
For a qualitative analysis of CN and CH abundances, we measured the modified 
S3839 and CH4300 indices used by 
\citet{H/S/G:03}, which are defined as
\begin{equation}
{\rm S3839= -2.5\;log \frac{F_{3861-3884}}{F_{3894-3910}}},
\end{equation}
\begin{equation}
{\rm CH4300= -2.5\;log \frac{F_{4285-4315}}{0.5F_{4240-4280}+0.5F_{4390-4460}}}
\end{equation}
where F are the fluxes in the different bandpass regions.
For comparison, we again measured the same indices in the 
M55 stars, expecting to show no strong variations in the light-element abundances.

In the following we describe the correction of the measured indices for
effects of temperature and compare the relative abundances of the thus-weighted indices for the different sub-populations in \wcen.
An absolute calibration of $\alpha$ and C, N abundances is foreseen for a future study.

\subsection{Correction of temperature effects}
We restricted our analysis of the light elements to the cooler/redder part of 
the subgiant branch, in the magnitude and colour range $16.9<V_0<17.8$ and
$(B-V)_0>0.44$. Since even this colour-selected set of stars shows a 
considerable range in temperature, we need to apply a correction that 
eliminates the effects of temperature on the strength of the absorption
lines.

\begin{figure}
\psfig{figure=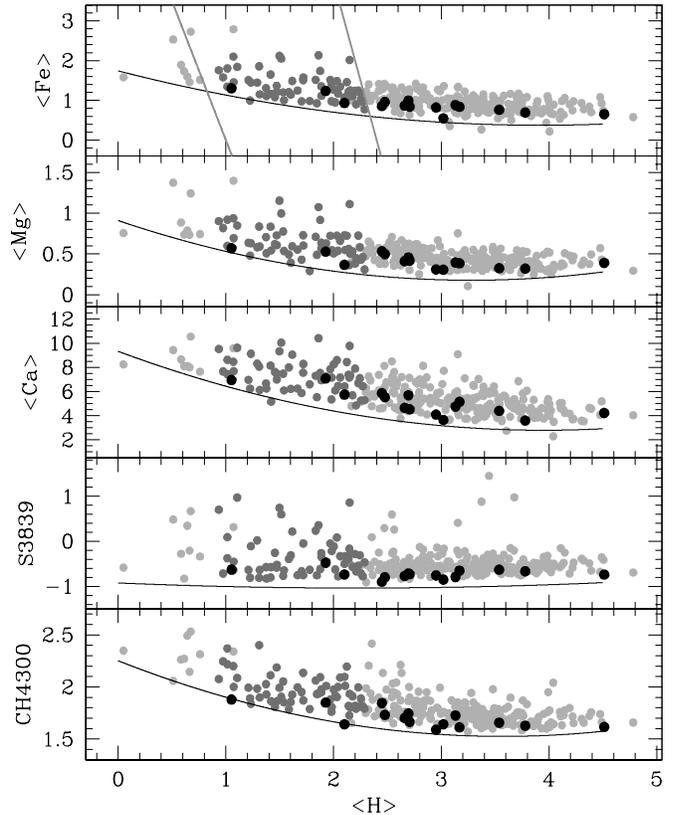,height=11.0cm,width=8.6cm,bbllx=9mm,bblly=60mm,bburx=158mm,bbury=246mm}
\caption{\label{cn1} Different line indices versus the hydrogen index \mh .
From top to bottom: iron indicator \mfe , magnesium index 
$\langle{\rm Mg}\rangle$, calcium index $\langle{\rm Ca}\rangle$, CN-band
indicator S3839, and CH-band index CH4300. The black dots are stars from the
comparison cluster M55. Dark grey dots indicate those stars that were selected
for further analyses. Besides the boundaries in
the upper panel, the selection criteria are a colour range of 0.5$<(B-V)_0<$0.68 mag.}
\end{figure}

As a temperature indicator, we used the combined hydrogen
index \mh\ (see Sect.~\ref{effects}). Figure~\ref{cn1} shows the dependence of 
the different element indices on \mh . As expected, the line strengths for the
different elements/molecules increase on average with decreasing \mh\ 
(decreasing temperature). For CN, however, there is no such strong dependence since the strength of the CN line at 3839 \AA\ only weakly depends on temperature \citep[see e.g.][]{J/J:95}.

In order to correct for the effects of temperature, we 
fitted a second order polynomial to the lower boundary of each index in 
Fig.~\ref{cn1}. The `temperature corrected' index is then calculated from the 
vertical distance $\langle{\rm index}\rangle-$fit, and the resulting value (`$\delta$') is further scaled by the median value for a given index. In the following, these corrected and scaled indices are denoted by, e.g., $\overline{\delta\langle{\rm Mg}\rangle}$.

Due to the stronger line indices at lower temperatures, the abundance 
variations can be analysed best in the regime of low \mh\ values. For the 
subsequent analyses, we therefore selected SGB stars between $1<$\mh$<2.2$. The 
exact selection boundaries are shown in Fig.~\ref{cn1}. Their inclinations are 
motivated by lines of equal temperatures as derived from the synthetic spectra 
in Fig.~\ref{calib}. The selected temperature range is about 5300-6100 K. 
Furthermore, we restricted the sample to the colour range $0.50<(B-V)_0<0.68$ 
mag. All these selection 
criteria ensure a minimisation of the temperature effects while maximising the 
signal of intrinsic abundance variations between the individual SGB stars.
It should be further noted that the restriction to stars on the subgiant branch
mostly excludes possible effects of atmospheric mixing as would be expected to 
occur in giant branch stars \citep[see e.g.][]{T/E:04}.

\begin{figure}
\psfig{figure=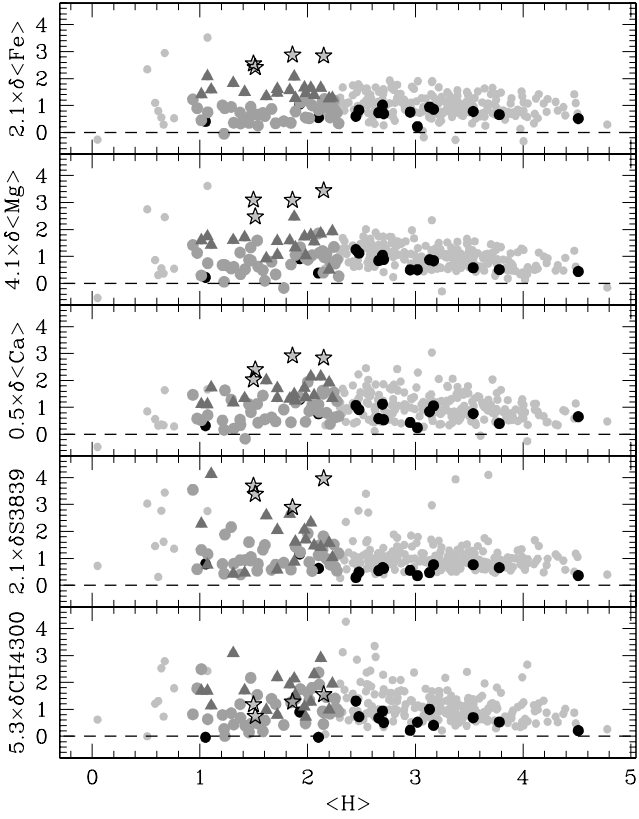,height=11.0cm,width=8.6cm,bbllx=9mm,bblly=60mm,bburx=158mm,bbury=246mm}
\caption{\label{cn2} Normalised and `temperature-corrected' line indices versus
the hydrogen index \mh . The black dots are stars from M55. Light dots are all the
data points for \wcen . The darker symbols mark the selected stars (see
Fig.~\ref{cn1}) in different metallicity bins: dots indicate stars with iron 
abundances between $-2.0$ and $-1.4\unit{dex}$; the grey triangles are for 
those with $-1.4<$[Fe/H]$<-1.0\unit{dex}$, and the dark asterisks are for stars
with $-1.0<$[Fe/H]$<-0.5\unit{dex}$.}
\end{figure}

\begin{figure}
\psfig{figure=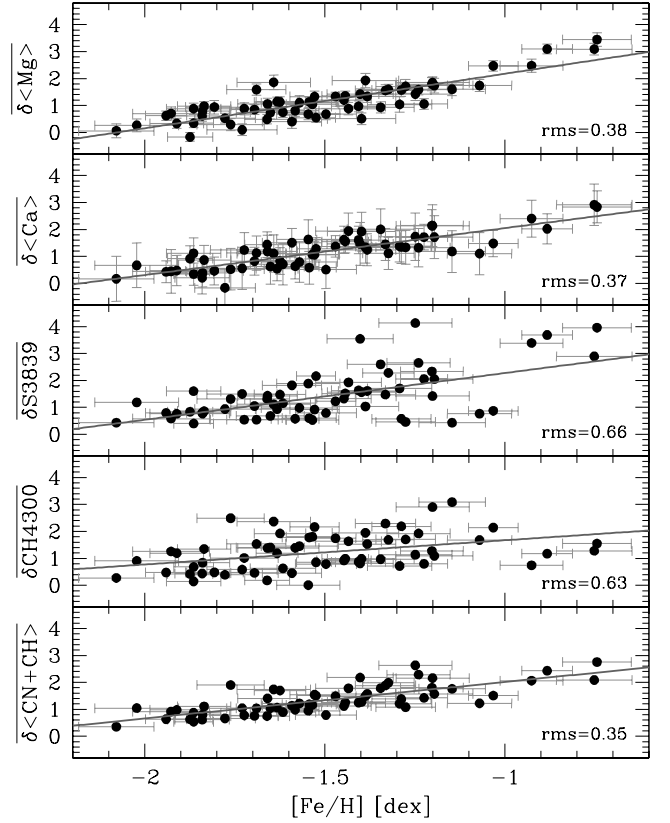,height=11.0cm,width=8.6cm,bbllx=9mm,bblly=60mm,bburx=158mm,bbury=246mm}
\caption{\label{cn3} Normalised and `temperature-corrected' line indices versus
iron abundance [Fe/H]. Only stars in the selected regions of Figs.~\ref{cn1} and
\ref{cn2} are considered. The lines are fits to the data points with the rms
indicated. Clearly, the scatter of the combined index \dcnch\ is significantly 
smaller than those of the individual indices alone.}
\end{figure}

\subsection{The dependencies of light elements on iron abundances}
The normalised and temperature-corrected indices as a function of the metallicity \feh\ 
are shown in Fig.~\ref{cn3}. 
It can be seen that for the five indices analysed (\dmg , \dca , \dch , \dcn , and \dcnch) 
there is a correlation in the sense that the line
strengths increase for increasing metallicities. Given that we only consider
stars on the SGB with (on average) the same luminosity and since we only make
use of temperature corrected line indices, we conclude that
the observed dependencies reflect the chemical enrichment history of these
stars in $\omega$ Cen. Interestingly,  the most metal-rich stars (\feh\ $\geq$ $-$1 dex) do not show a significantly larger \dch\ as compared to e.g. stars with \feh\ $\sim$ $-$1.5 dex.
For \dcn , however, we find significantly higher values for the metal-rich population (but note the two CN-rich stars at \feh\ $\sim$ $-$1.4 dex).
In particular, there is an anticorrelation among \dcn\ and \dch\ for the most metal-rich stars. This could most readily be explained by the enrichment of CNO-processed material (with increased N and decreased C).
However, that only the most metal-rich stars seem to show this anticorrelation is puzzling.
A careful examination of the location of these stars in the colour-magnitude diagram shows that the most metal-rich stars are not the reddest ones. Moreover, these stars have rather intermediate luminosities (as compared to the overall sample of stars) at these colours. This indicates that the observed anticorrelation is  not due to (not corrected for) temperature and/or luminosity effects or that these stars occupy a special region in the colour-magnitude diagram. Instead, it could be that the enrichment history for these stars was indeed very different from that of the earlier populations.
The scatter as measured by
the rms of the best-fitting line is much larger for \dcn\ and \dch\
than for the $\alpha$ elements Mg and Ca (as measured by
\dmg\ and \dca ). Under the assumption that \dcn\ mostly depends
on N while  \dch\ is a measure of C abundance, we conclude that there is indeed a wide range of C and N abundances, similar to
what was found for RGB stars by \citet{N/DC:95}.

It should be also noted that the average index $(\delta S3839+\delta CH4300)/2$
($=$ \dcnch\ in Fig.~\ref{cn3}) shows a strong correlation with increasing 
\feh\ and a significantly smaller scatter as compared to the individual indices \dcn\ 
and \dch . This was also observed by \citet{H/R:00}, albeit
for RGB stars. 
The fact that the CN/CH anticorrelation is not only visible on the RGB but is also detected for almost  unevolved stars on the SGB strongly suggests that mixing effects cannot be the dominant cause of the increase of C and N for higher metallicities.
Instead, this shows that the enrichment of subsequent stellar populations by the light
elements followed that of the iron peak elements.

\subsection{Line strengths of $\alpha$ elements}\label{secalpha}
\begin{figure}
\psfig{figure=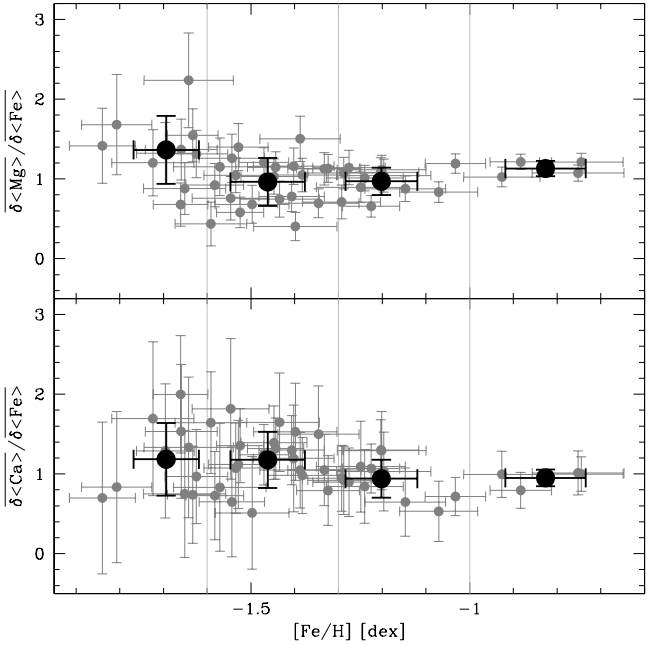,height=8.6cm,width=8.6cm,bbllx=9mm,bblly=60mm,bburx=195mm,bbury=246mm}
\caption{\label{cn4} Ratio of magnesium (upper panel) and calcium (lower panel)
to the iron index versus iron abundance [Fe/H] for selected stars (see text). The
bold data points are averages for the given metallicity bins (vertical lines).}
\end{figure}

The $\alpha$-elements are important tracers for the enrichment by SN\,II events.
Figure~\ref{cn4} shows the temperature-corrected and 
scaled Mg over Fe and Ca over Fe absorption-line strengths as a function of 
\feh . For both elements, we find that the ratio remains almost constant over 
the whole metallicity range (up to $\sim$ $-$0.8 dex). 
The flatness of the $\alpha$ elements, here for SGB-stars, has also been found by
\citet{N/DC:95} for RGB-stars in the same metallicity range.
This is interpreted as a signature for an enrichment dominated by SN\,II events.
\citet{P/P:02} detected indications for SN\,Ia enrichment only at very high metallicity
($>-$0.8 dex), which our stars hardly reach.
Moreover, it should be noted that the number of stars at higher 
metallicities is rather low, thus lowering the statistical significance of
our analysis.

In general, it appears that the spread in [$\alpha$/Fe] is wider for lower
metallicities. This is possibly not only due to the larger number of
low-metallicity stars in our sample but rather because of the overall higher
uncertainty in the measurements for shallower absorption lines at overall lower
metallicities (as indicated by the larger error bars). Nevertheless, we cannot
exclude the possibility that some fraction of this scatter is due to the
chemical enrichment process in $\omega$ Cen. The scatter would then reflect
the enrichment by stochastical SN\,II events during the early star formation
phase.

\subsection{CN/CH line strengths}\label{seccnch}
\begin{figure}
\psfig{figure=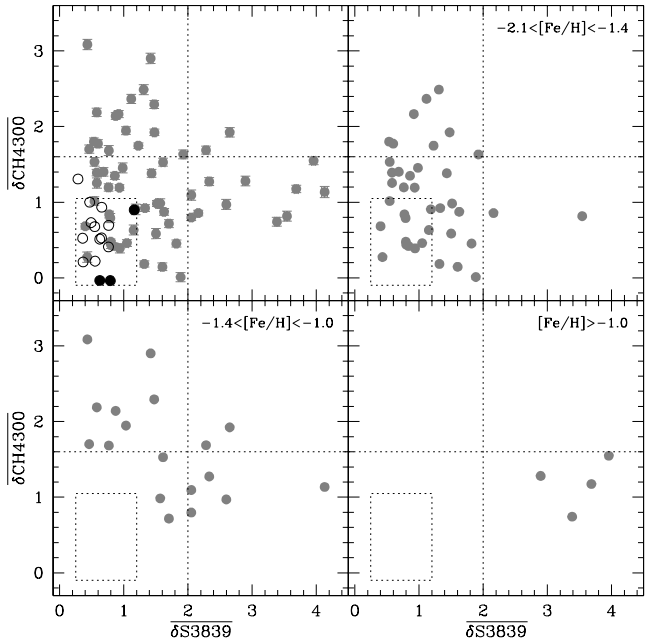,height=8.6cm,width=8.6cm,bbllx=9mm,bblly=60mm,bburx=195mm,bbury=246mm}
\caption{\label{cn5} Normalised and `temperature-corrected' CN versus CH index.
In the upper right panel all selected stars (see Fig.~\ref{cn1} and \ref{cn2})
are shown, together with the stars in M55 that are confined to a small area
(dotted box). Filled dots show stars in M55 in the selected temperature range, while open symbols show all examined stars in M55. The other panels show the CN versus CH index for three different
metallicity bins as indicated. The horizontal and vertical lines divide the group of
stars into CH-rich and CN-rich populations.}
\end{figure}

The dependence of the temperature-corrected, scaled indices for CN (\dcn ) on 
CH (\dch ) is shown in Fig.~\ref{cn5}. 
In general, we find that there is a large scatter in the CH and CN abundances 
of stars in $\omega$ Cen, as compared to e.g. M\,55. 
Indeed, while most of the stars in M\,55  are confined to a small
region in the CH vs. CN diagram, the stars in \wcen\ 
show a large scatter towards higher CH or higher CN abundances.

Such a spread was also found by \citet{S/DC:04} 
in their analysis of 450 MS and MSTO stars in $\omega$ Cen.
In their comparison with other galactic globular clusters, they
showed that this behaviour can be seen neither for any other globular cluster in
their sample nor for close Galactic disk and halo stars.
The only other globular cluster showing such a bimodal distribution of CN and CH
abundances, but to a much smaller degree, is 47 Tuc.
\citet{S/DC:04} argue that the reason for the unique CN/CH spread in \wcen\ might be the much larger iron abundance of its stars .

Figure~\ref{cn5} includes plots of \dch\ vs. \dcn\ for 
different ranges of the metallicity \feh . It can be seen that the 
low-metallicity stars show a wide spread in CH, but there are only few CN-rich 
stars. For high metallicities, however, there are only stars with moderate CH 
but with high CN-band strengths; i.e. these stars are possibly enriched in N and depleted in C.
The large scatter towards higher CH or  CN abundances
in the \dcn\ vs. \dch\ diagram (see also \citealt{S/DC:04}) could be explained
by a prolonged enrichment history in a galactic environment.
In a scenario where \wcen\ was the nucleus of a
gas-rich dwarf galaxy, the intermediate
metallicity sub-populations of \wcen\ would have been created from the enriched material of 
this underlying galactic disk population.

\section{Summary and conclusions}\label{summcon}
The spectra of 429 stars near the SGB/MSTO region of the globular cluster \wcen\
were analysed in the wavelength range $3500$ to $6200\unit{\AA}$.
The equivalent widths of newly-defined line indices were measured.
Absorption lines in the overlapping wavelength regions of the two grisms, as well as repeatedly observed stars, were used to prove the
consistency and to estimate the uncertainties of our measurements.
In order to calculate an index with high sensitivity to Fe, we combined seven
individual line indices to an average iron indicator \mfe .
Furthermore, we defined a mean Balmer index \mh\ that served as a temperature
indicator.

The same indices were measured on spectra of the chemically
homogeneous cluster M55, in addition to standard stars and synthetic spectra. These indices 
were used to estimate the dependence of \mfe\ on temperature and surface gravity
and to establish an analytical function [Fe/H]=f(\mfe, \mh). The error in the
absolute iron abundance calibration is of the order $\pm$ 0.1 to 0.3 dex.

Furthermore, we measured Mg, Ca, CN, and CH indices for the \wcen\ and M55
spectra to study the abundance variations in the different populations.
The [$\alpha$/Fe] ratio is fairly flat over a wide range of metallicities, 
indicating a prolonged enrichment by massive (M $>$ 10 \Msun) stars.
Moreover, we showed that the strong CN and CH variations
(which had been found on the RGB before, see e.g. \citealt{H/R:00}) are also found among the stars in the MSTO/SBG region, 
i.e. these variations most possibly have a primordial origin so are not due to mixing effects.
While the combined CN+CH abundance smoothly increases with
metallicity, the enrichment in C and N is anti-correlated. The most metal-rich
stars show an extreme enrichment in the CN-bands, whereas metal-poor stars
are scattered to high CH abundances, probably pointing to a fast C enrichment
by more massive AGB stars. 

The abundance patterns of the stars in \wcen\ are uncommon among the globular
cluster population of our Milky Way. This refers not only to the iron abundance but also to the lighter element abundances.
The strong abundance variations of the light elements indicate that \wcen\ 
experienced a very complex enrichment 
history and that the progenitor of this object must have been a far more 
massive object that was able to retain ejecta by its subsequent
generations of stars. Whether this enrichment was due to homogeneous or 
episodic star formation must be left to a more detailed analysis that includes 
information on s- and r-process elements, which would allow more accurate
models of the star formation rate and the initial stellar mass distributions to be derived.

Over the past years, many studies have revealed more and more
peculiarities of \wcen . 
Deep photometry enabled a more detailed inquiry of the bifurcation of the
MS found by \citet{A:97}, which can most probably be explained as an He
overabundance of the intermediate metallicity population (\citealt{N:04} and 
\citealt{P/V:05}), thus complicating the interpretation of the chemical-enrichment history of \wcen .
Up to now there no self-consistent models exist for the origin of \wcen ,
but most studies suggest a complex formation scenario in order to explain the unusual observed properties of this  globular cluster.
Our study provides further evidence of a scenario in which \wcen\ was 
embedded in a formerly larger galactic entity that was captured and 
disrupted by the Milky Way.

In our previous study \citep{H/K:04}, we show that \wcen\ has experienced
an extended star-formation history over a period of at least 3 Gyr.
In combination with the abundance patterns presented in this study, the combined 
information shares many similarities with what is found for nearby dwarf 
galaxies that are known to have experienced rather complex star-formation 
processes \citep[e.g.][]{S/V:03, T/V:02, L/M:03}.

In general, star formation histories vary from dwarf to dwarf galaxy. Both a 
continuous star formation and bursts seem to be common among the Local Group dSphs \citep{G:99}.
Furthermore, \citet{H/G:01} have found significant population gradients in
Local Group dSphs in the sense that the more metal-rich and younger 
sub-populations are concentrated towards the galaxies' centres.
This is also observed among the stars in  \wcen\  \citep[e.g.][]{N/F/M:96},
thus in agreement with a galactic origin of this cluster.

In a forthcoming paper a much more detailed analysis of a larger sample of 
stars in $\omega$ Cen is foreseen. Combined with precise photometry from HST,
the larger spectroscopic data set will be used to shed more light on the 
enrichment history of this cluster. In particular, we will take the suggested 
variations in He abundances into account, which can have a non-negligible effect
on the derived ages for the different stellar populations.

\acknowledgements
The authors are very grateful to S.-C. Rey  for providing the $BV$ data. We 
also thank T. Puzia for providing gonzo, and thanks to the anonymous referee for the very helpful comments.\\


\bibliographystyle{aa}
\bibliography{kayser}


\enddocument